\documentclass[12pt]{article}

\usepackage{a4}

\begin{document}
\title{Winding number versus Chern--Pontryagin charge}
\author{{\large Tigran Tchrakian}$^{\dagger \star}$ \\ \\
$^{\dagger}${\small Department of
Mathematical Physics, National University of Ireland Maynooth,} \\
{\small Maynooth, Ireland} \\ \\
$^{\star}${\small Theory Division, Yerevan Physics Institute (YerPhI),
AM-375 036 Yerevan 36, Armenia}}

\date{}
\newcommand{\dd}{\mbox{d}}
\newcommand{\tr}{\mbox{tr}}
\newcommand{\la}{\lambda}
\newcommand{\ka}{\kappa}
\newcommand{\al}{\alpha}
\newcommand{\ga}{\gamma}
\newcommand{\de}{\delta}
\newcommand{\si}{\sigma}
\newcommand{\bomega}{\mbox{\boldmath $\omega$}}
\newcommand{\bsi}{\mbox{\boldmath $\sigma$}}
\newcommand{\bchi}{\mbox{\boldmath $\chi$}}
\newcommand{\bal}{\mbox{\boldmath $\alpha$}}
\newcommand{\bpsi}{\mbox{\boldmath $\psi$}}
\newcommand{\brho}{\mbox{\boldmath $\varrho$}}
\newcommand{\beps}{\mbox{\boldmath $\varepsilon$}}
\newcommand{\bxi}{\mbox{\boldmath $\xi$}}
\newcommand{\bbeta}{\mbox{\boldmath $\beta$}}
\newcommand{\ee}{\end{equation}}
\newcommand{\eea}{\end{eqnarray}}
\newcommand{\be}{\begin{equation}}
\newcommand{\bea}{\begin{eqnarray}}
\newcommand{\ii}{\mbox{i}}
\newcommand{\e}{\mbox{e}}
\newcommand{\pa}{\partial}
\newcommand{\Om}{\Omega}
\newcommand{\vep}{\varepsilon}
\newcommand{\bfph}{{\bf \phi}}
\newcommand{\lm}{\lambda}
\def\theequation{\arabic{equation}}
\renewcommand{\thefootnote}{\fnsymbol{footnote}}
\newcommand{\re}[1]{(\ref{#1})}
\newcommand{\R}{{\rm I \hspace{-0.52ex} R}}
\newcommand{\N}{{\sf N\hspace*{-1.0ex}\rule{0.15ex}%
{1.3ex}\hspace*{1.0ex}}}
\newcommand{\Q}{{\sf Q\hspace*{-1.1ex}\rule{0.15ex}%
{1.5ex}\hspace*{1.1ex}}}
\newcommand{\C}{{\sf C\hspace*{-0.9ex}\rule{0.15ex}%
{1.3ex}\hspace*{0.9ex}}}
\newcommand{\eins}{1\hspace{-0.56ex}{\rm I}}
\renewcommand{\thefootnote}{\arabic{footnote}}

\maketitle
\begin{abstract}
In the {\it usual} $d$ dimensional $SO(d)$ gauged Higgs models with
$d$-component Higgs fields, the 'energies' of the topologically stable
solitons are bounded from below by the Chern-Pontryagin charges. A new
class of Higgs models is proposed here, whose 'energies' are stabilised
instead by the {\it winding number} of the Higgs field itself, with no
reference to the gauge group. Consequently, such Higgs models can be
gauged by $SO(N)$, with $2\le N\le d$.
\end{abstract}
\medskip
\medskip

\section{Introduction}
\label{introduction}
The $d$ (Euclidean) dimensional $SO(d)$ Higgs
models~\cite{models2,models3,models4}
that result from the dimensional descent of the generalised
Yang-Mills~\cite{GYM} (GYM) model on $\R^d\times S_{4p-d}$, support
soliton/instanton solutions (topologically) stabilised by the
residual $2p$-th Chern-Pontryagin (C-P) charge\cite{charges} resulting
from the dimensional reduction of $2p$-th C-P charge on
$\R^d\times S_{4p-d}$. These residual C-P charges are surface
integrals~\cite{charges} of Chern-Simons (C-S) densities depending
on the $SO(d)$ connection and Higgs fields $(A_i^{[ab]},\phi^a)$,
with $i=1,2,..,d$ and $a=1,2,..,d$.

In the title, the term Chern-Pontryagin charge means the residual $2p$-th
C-P charge in $d$ dimensions, in the model descending from the GYM system
on $\R^d\times S_{4p-d}$. A prominent property of these charges is that
the C-S densities whose surface integrals they are, are {\em gauge
invariant} when $d$ is odd, and are {\em gauge variant} when $d$ is even.
For example for $d=3$, the model coincides with the Georgi-Glashow model
in the Prasad-Sommerfield limit and the residual C-S density is the
{\em gauge invariant} quantity
\be
\label{mon}
\Omega_i^{(3)}=\frac{1}{8\pi}\ \vep_{ijk}\vep^{abc}\phi^a F_{jk}^{[ab]}
\ee
which is the magnetic field strength at infinity, while for $d=2$ and
$d=4$ these are respectively
\bea
\Omega_i^{(2)}&=&\frac{1}{2\pi}\ \vep_{ij}\left(A_j
+\vep^{ab}\phi^aD_j\phi^b\right)\label{vor}\\
\Omega_i^{(4)}&=&\frac{1}{8\pi^2}\ \vep_{ijkl}\vep^{abcd}A_j^{[ab]}\left(
F_{kl}^{[ab]}+\frac23\left(A_kA_l\right)^{[ab]}\right)
+{\rm Higgs \ dependent \ terms}\label{inst}\ .
\eea
The 'surface@ integral of \re{vor} is the winding number of the
Abrikosov-Nielsen-Olesen vortex with the dynamics of the residual Higgs
model given by the Abelian Higgs model. In \re{inst} we have omitted
the Higgs dependent terms because we do not wish to refer in detail
to the rather exotic model~\cite{models2,models3,models4} describing the
dynamics, but
note that in both \re{vor} and \re{inst} the leading {\em gauge variant}
term coincides with the C-S density of the purely gauge field system.

Our purpose here is to eschew the application of the C-P charges
\re{mon}-\re{inst} as the topological charges, in favour of employing
instead the winding numbers of the Higgs fields. Indeed the only reason
for displaying \re{mon}-\re{inst} is to state our particular nomenclature
for reduced C-P and C-S densities in the context of the Higgs model
considered here. Also, we will have occasion to refer to \re{vor} in
detail, because it will turn out that the C-P charge calculated from
\re{vor} coincides with the topological charge when we instead stabilise
the soliton witt the winding number.

The definition of the winding number of the $d$ component Higgs field
$\phi^a$ in $d$ dimensions is
\be
\label{wind}
N=\int\varrho_0\ d^dx=
\frac{1}{\Omega_d}\int\vep_{i_1i_2..i_d}\vep^{a_1a_2..a_d}
\pa_{i_1}\phi^{a_1}\pa_{i_2}\phi^{a_2}..\pa_{i_d}\phi^{a_d}\ d^dx
\ee
where $\Omega_d$ is the angular volume. It is implicit in our thinking
here that the Higgs field obeys the usual asymptotics
\be
\label{higgs}
\lim_{r\rightarrow\infty}|\phi^a|^2=\eta^2\ ,
\ee
where $\eta$ is the VEV. Since $\varrho_0$ defined by \re{wind} is a
total divergence, \re{wind} can be
evaluated as a surface integral. But unlike the {\em gauge invariant}
C-P densities pertaining to \re{mon}-\re{inst}, \re{wind} is manifestly
{\em gauge variant}. This property disqualifies its direct application
as a topological charge presenting a lower bound to a gauge theory --
in which case the charge must be {\em gauge invariant}.

As the title suggests however, it is the winding number \re{wind} which
we propose as an alternative to the C-P charge notwithstanding the
{\em gauge variance} of the former. What has to be done is to find a
{\em gauge invariant} topological charge density (versus the C-P density)
whose volume integral is equal to \re{wind}\footnote{In the case of $d$
dimensional $SO(d)$ gauged $O(d+1)$ sigma models~\cite{LMP} whose
dynamics leaves the gauge group unbroken the C-P charges vanish so that
the only topological charge available is of this type.}.

In section {\bf 2} the {\em gauge invariant} topological charges that
are the alternatives to the C-P charges will be constructed. In section
{\bf 3} the topological lower bounds will be established for the
simplest models, which in all $d$ except $d=2$ differ from the
corresponding models whose solitons are stabilised by the C-P charges.
In both sections {\bf 2} and {\bf 3} we restrict our
considerations to the $d=2$ and $d=3$ cases, mainly to avoid the
complications of dealing with the more exotic Higgs
models~\cite{models2,models3,models4}
in $d\ge 4$. In section {\bf 4} we discuss the results and point out
some applications of the new models.

\section{Topological charge density}
\label{topch}
The winding number \re{wind} is not gauge invariant. One could replace
the integrand in \re{wind} by the gauge invariant density
\be
\label{g}
\varrho_G=
\frac{1}{\Omega_d}\vep_{i_1i_2..i_d}\vep^{a_1a_2..a_d}
D_{i_1}\phi^{a_1}D_{i_2}\phi^{a_2}..D_{i_d}\phi^{a_d}\ ,
\ee
but unlike \re{wind}, the volume integral of this density cannot be
evaluated as a surface integral and hence cannot be a candidate for
a toplogical charge which can be evaluated using only the asymptotic
values of the fields. This is because \re{g} is not a total divergence.

As we stated above, our aim is to find a new density which is gauge
invariant and whose volume integral equals the winding number \re{wind},
i.e.  that it is a topological charge. As in the case of sigma
models~\cite{LMP}, we express the density $varrho_G$ \re{g} in terms of
$varrho_0$ (see\re{wind}). Then using the Leibniz rule for covariant
derivatives we split the difference $(varrho_G-\varrho_0$ into a
gauge variant total plus a gauge invariant terms.

This procedure can be carried out for any $d$ but we will restrict here
to the $d=2$ and $d=3$ cases only, which are given in the next two
subsections.

\subsection{$SO(2)$ gauged charge in $d=2$}
In this case
\be
\label{rho2}
\varrho_0=\frac{1}{2\pi}\vep_{ij}\vep^{ab}\pa_i\phi^a\pa_j\phi^b\quad ,
\quad\varrho_G=\frac{1}{2\pi}\vep_{ij}\vep^{ab}D_i\phi^aD_j\phi^b
\ee
where $D_i\phi^a$, also appearing in \re{vor}, is defined as
\[
D_i\phi^a=\pa_i\phi^a+A_i\vep^{ab}\phi^b\ .
\]
Then we can write the difference of the densities in \re{rho2}, using the
Leibniz rule, as
\be
\label{diff2}
\varrho_G-\varrho_0=-\frac{1}{2\pi}\vep_{ij}\left(\pa_i(|\phi^a|^2A_j)
-\frac12|\phi^a|^2F_{ij}\right)\ .
\ee

It is now necessary for our purposes to have a vanishing contribution
from the volume integral of the total divergence term in \re{diff2},
which as it stands does not vanish by virtue of \re{higgs}. this can be
rectified by adding and subtracting one half $\eta^2$ times the first C-P
density\footnote{This use of the C-P density is a general feature in all
even dimensions\cite{LMP}.}
$\vep_{ij}F_{ij}$ to the right hand side of \re{diff2}. The
result is
\be
\label{diff2a}
\varrho_G-\varrho_0=-\frac{1}{2\pi}\vep_{ij}
\left(\pa_i[(\eta^2-|\phi^a|^2)A_j]
-\frac12(\eta^2-|\phi^a|^2)F_{ij}\right)\ ,
\ee
the volume integral of which receives contributions only from the gauge
invariant term on the right hand side of \re{diff2a}. It is natural then
to define a topological charge
\bea
\varrho&=&\frac{1}{2\pi}\left(\varrho_0+
\vep_{ij}\pa_i[(\eta^2-|\phi^a|^2)A_j]\right)\label{top2v}\\
&=&\frac{1}{2\pi}\left(\varrho_G
+\frac12\vep_{ij}(\eta^2-|\phi^a|^2)F_{ij}\right)\ .\label{top2i}
\eea
From \re{top2v} it is obvious that the volume integral of $\varrho$ is
just the winding number \re{wind}, while at the same time it is obvious
from \re{top2i} that $\varrho$ is gauge invariant.

It is noteworthy that \re{top2i} can be rewritten as
\be
\label{vor2}
\varrho=\frac{1}{2\pi}\vep_{ij}\pa_i\left(A_j
+\vep^{ab}\phi^aD_j\phi^b\right)\ ,
\ee
which yields precisely the C-S density \re{vor}. Thus the new topological
charge we have defined is identical to the C-P charge. This is a low
dimensional accident.

\subsection{$SO(3)$ gauged charge in $d=3$}
In this case
\be
\label{rho3}
\varrho_0=\frac{1}{4\pi}\vep_{ijk}\vep^{abc}
\pa_i\phi^a\pa_j\phi^b\pa_k\phi^c\quad ,
\quad\varrho_G=\frac{1}{4\pi}\vep_{ijk}\vep^{abc}
D_i\phi^aD_j\phi^bD_k\phi^c
\ee
where $D_i\phi^a$, also appearing in \re{mon}, is defined as
\[
D_i\phi^a=\pa_i\phi^a+A_i^{[ab]}\phi^b\ .
\]
Then we can write the difference of the densities in \re{rho2}, using the
Leibniz rule, as
\be
\label{diff3}
\varrho_G-\varrho_0=\frac{1}{4\pi}.\frac23\vep_{ijk}\vep^{baa'}\left[
\pa_i\left(A_j^{aa'}\phi^b\pa_k|\phi^c|^2\right)-
F_{ij}^{aa'}\phi^b\pa_k|\phi^c|^2\right]\ .
\ee

The volume integral of the total divergence term in \re{diff3} now
vanishes outright. This is a feature in all odd dimensions.(Recall that
in the $d=2$ case it did not.) To see this consider it as a surface
integral
\be
\label{surf}
I=\frac{1}{4\pi}.\frac23\int\vep_{ijk}\vep^{baa'}
A_j^{aa'}\phi^b\pa_k|\phi^c|^2dS_i .
\ee

For $I$ to be nonvanishing the integrand in \re{surf} must decay
asymptotically as $r^{-2}$ {\em and no faster}. Now the connection
$A_j^{aa'}$ decays as $r^{-1}$ by {\em finite energy conditions}, while
\re{higgs}, which enforces the asymptotic constancy of $|\phi^c|^2$
implies that $\pa_k|\phi^c|^2$ decays as $r^{-(2+\epsilon)}$,
$\epsilon>0$. Thus the integrand in \re{surf} decays as
$r^{-(3+\epsilon)}$ and hence $I=0$.

It is natural now to define the (gauge invariant) topological charge
density as
\bea
\varrho&=&\varrho_0+\frac{1}{4\pi}.\frac32\vep_{ijk}\vep^{baa'}
\pa_i\left(A_j^{aa'}\phi^b\pa_k|\phi^c|^2\right)\label{top3v}\\ &=&
\varrho_G+\frac{1}{4\pi}.\frac32\vep_{ijk}\vep^{baa'}
F_{ij}^{aa'}\phi^b\pa_k|\phi^c|^2\label{top3i}\ .
\eea
Again it is obvious from \re{top3v} and \re{surf} that the volume integral
of this charge equals the winding number \re{wind} for $d=3$, and from
\re{top3i} it is obvious that this is a gauge invariant charge density.

The topological charge density \re{top3i} just defined differs from the
corresponding C-P density \re{mon}. Thus the identity of the new
topological charge density (whose volume integral equals the winding
number) with the corresponding C-P density occurs only for $d=2$. The new
topological charge, namely the winding number, is distinct from the
corresponding C-P charge for all $d\ge 3$.

\section{Lower bounds: Gauged Higgs models}
\label{l-bounds}
Having constructed the required topological charge densities in the
previous section, we proceed to define new models whose energies/actions
are bounded from below by the winding number. These lower bounds are
established employing Bogomol'nyi type inequalities. This can be done for
the case of any $d$, but as in the previous section we will concern
ourselves with the cases $d=2$ and $d=3$ only. These are given in the
next two subsections.

\subsection{$SO(2)$ gauged model in $d=2$}
Since the new topological charge density \re{top2i} in this case coincides
with the C-P charge, it follows that the pertinent Bogomol'nyi type
inequalities for this case are the original Bogomol'nyi inequalities for
which we refer to Ref.~\cite{bog}. The corresponding Hihhs model is of
course the familiar Abelian Higgs model.

\subsection{$SO(3)$ gauged model in $d=3$}
As the new topological charge density \re{top3i} in this case differs
from the C-P charge obtained from the surface integral of \re{mon}, the
corresponding topological (Bogomol'nyi) inequalities and thence the Higgs
models that follow, differ (very appreciably) from the Georgi-Glashow
model in the Prasad-Sommerfield limit.

The pertinent Bogomol'nyi inequalities are
\bea
\left|D_i\phi^a-\frac14\ka_1^2\vep_{ijk}\vep^{abc}
D_{[j}\phi^bD_{k]}\phi^c\right|^2&\ge&0\label{bog1}\\
\left|F_{ij}^{aa'}-\frac34\ka_2^2\vep_{ijk}\vep^{aa'b}
\phi^b\pa_k|\phi^c|^2\right|^2&\ge&0\label{bog2}\ ,
\eea
where $\ka_1$ and $\ka_2$ are constants with dimensions $[L^2]$, given
that the Higgs field has dimension $[L^{-1}]$. Expanding \re{bog1} and
\re{bog2} yields, respectively
\bea
\frac12|D_i\phi^a|^2+\frac18\ka_1^4|D_{[i}\phi^aD_{j]}\phi^b|^2&\ge&
\frac12\ka_1^2\ .\ 4\pi\varrho_G\label{bog1.2}\\
\frac18|F_{ij}^{ab}|^2+\frac{9}{32}\ka_2^4|\phi^a|^2
|(\pa_i|\phi^b|^2)|^2&\ge&\frac{3}{16}\ka_2^2\vep_{ijk}\vep^{aa'b}
F_{ij}^{aa'}\phi^b(\pa_k|\phi^c|^2)\label{bog2.2}\ .
\eea
Adding \re{bog1.2} and \re{bog2.2} yields the final Bogomol'nyi bound
for the energy density, i.e. the left hand side of this inequality defines
the energy density functional (the static Hamiltonian) and the right hand
side coincides with $2\pi\ka_1$ times the topological charge density
\re{top3i}, provided that the constants $\ka_1$ and $\ka_2$ satisfy the
following condition is satisfied
\be
\label{condn}
\ka_2=2\ka_1\ .
\ee
Thus, denoting $\ka_1^4=\ka$ we finally have
\be
\label{H}
{\cal H}\stackrel{\rm def}=\frac18|F_{ij}^{ab}|^2+\frac12|D_i\phi^a|^2+
\frac18\ka\left(|D_{[i}\phi^aD_{j]}\phi^b|^2+36|\phi^a|^2
|(\pa_i|\phi^b|^2)|^2\right)\ge 2\pi\varrho\ ,
\ee
with $\varrho$ given by \re{top3i} (and `\re{top3v}) so that its volume
integral is equal to the winding number of the Higgs field at infinity.

The normalisations in \re{H} are chosen so that when $\ka=0$ the left hand
side reduces to the Georgi-Glashow system in the Prasad Sommerfield limit,
in which case we have the usual lower bound
\bea
{\cal H}_{PS}&=&\frac18|F_{ij}^{ab}|^2+\frac12|D_i\phi^a|^2\ge
4\pi\varrho_M\label{PS}\\
\varrho_M&=&\frac{1}{4\pi}\vep_{ijk}\vep^{aa'b}\pa_k\ (F_{ij}^{aa'}\phi^b)
\label{monch}
\eea
where the volume integral of $\varrho_M$ defined by \re{monch} equals the
C-P (monopole) charge $\mu$ exactly. Obviously $\cal H$ defined by \re{H},
in addition to being bounded from below by $2\pi\varrho$, is also bounded
by $4\pi\varrho_M$.

While the inequality \re{PS} is saturated by
\be
\label{self}
F_{ij}^{ab}=\vep_{ijk}\vep^{abc}D_k\phi^ac\ ,
\ee
the inequality \re{H} cannot be saturated since that would be tantamount
to saturating the two inequalities \re{bog1} and \re{bog2}, which is an
overdetermined system.

In both systems \re{H} and \re{PS} we have assumed that the Higgs field
obeys the asymptotic condition \re{higgs}. Otherwise the expected
topological lower bounds would not be valid. This is, as usual, ensured
dynamically by adding the usual Higgs potential
\be
\label{pot}
V(|\phi^a|)=\la(\eta^2-|\phi^a|^2)^2\ ,
\ee
in which case the C-P inequality corresponding to \re{PS} cannot be
saturated too.

\section{Discussion and outlook}
\label{disc}
In the previous two sections we have seen the construction of gauge
invariant topological charge densities which present lower bounds on
the energy densities of the corresponding $SO(d)$ gauged Higgs models,
and whose volume integrals equal the winding number \re{wind}. This was
done for dimensions $d=2,3$ but can systematically be extended to
arbitrary $d$.

A noteworthy feature of these examples is that in the $d=2$ case the
topological charge density thus defined happens to coincide with the
C-P density. It is clear from the $d=3$ example however, that for all
$d\ge 3$ this new charge density does {\bf not} coincide with the C-P
density.

There are two interesting applications which arise from the present
formulation. The first is dynamical. As we see from the work of section
{\bf 3.2}, the energy density \re{H} involves a quartic kinetic (Skyrme)
term multiplying the coupling constant $\ka$. But we know from our
experience with such theories~\cite{KS,KOT,BHT} that these support
attractive like-charged solitons, namely Skyrmions~\cite{S} in \cite{KS},
monopoles in \cite{KOT} and sigma-model monopoles in \cite{BHT}. Thus, we
would expect that the theory described by \re{H} also supports attractive
like-charged monopoles. If one included the Higgs potential \re{pot} in
\re{H}, this would render the decay of the Higgs field exponential and
hence would allow the repulsive Coulomb effect of the gauge field to
dominate. But as long as there is a Skyrme term present, this effect
is expected to be cancelled by the attractive effect of the latter, as
was found in \cite{KOT}.

The second application hinges on the fundamental property of the new
(gauge invariant) topological charge density, e.g. \re{top2v} and
\re{top3v}, namely that because the gauge field dependent surface
integrals vanish, its volume integral equals \re{wind} and is independent 
of the gauge connection and hence the gauge group. This indicates that
the gauge group can be taken to be $SO(N)$ with $2\le N\le d$ so that
only $N$ of the $d$ components of the Higgs field are gauged. In $d=3$ for
example this would enable the construction of a $U(1)$ soliton. This
would be the Higgs model analogue to the $U(1)$ gauged Skyrmion
constructed in \cite{PT} and like the latter would necessarily be axially
symmetric and would have zero monopole flux and a nonvanishing
magnetic dipole moment.

Unfortunately there is a price to pay in the
construction of such a $U(1)$ gauged Higgs theory whose energy is bounded
from below by the winding number, namely the volume integral of the first
member of \re{rho3}. The reason for this is dynamical. If the Higgs model
features the usual kinetic term
\be
\label{usual}
|D_i\phi^a|^2\ ,
\ee
then that model does not possess a gauge--decoupled limit, unlike the
corresponding Sigma model~\cite{PT}. Had a gauge--decoupled limit of the
model existed, we could surely have replaced the usual (Georgi-Glashow)
gauge group $SO(3)$ by $SO(2)$.

Let us demonstrate this fact and then explain how to ovecome this
obstacle, and spell out the attendent price to pay. Within the axially
symmetric Ansatz for the $U(1)$ field $A_i=(A_{\mu},A_3)\ ,\ \mu=1,2$
and the Higgs field $\phi^a=(\phi^{\al},\phi^3)\ ,\ \al=1,2$,
\be
\label{axial}
A_{\mu}=\frac{a(r,\theta)-N}{r\sin\theta}\vep_{\mu\nu}\hat x_{\nu}\ \ \ ,
\ \ \ A_3=0\ \ \ ,\ \ \ \phi^{\al}=\eta\ h(r,\theta)\ n^{\al}\\ \ ,
\quad\phi^3=\eta\ g(r,\theta)\ ,
\ee
with $n^{\al}=(\cos N\phi,\sin N\phi)$,
the reduced (two dimensional) quadratic kinetic term in \re{usual} reads
\be
\label{quadcov}
r^2|D_i\phi^a|^2=\eta^2\left(\sin\theta\left[r^2(h_r^2+g_r^2)+
(h_{\theta}^2+g_{\theta}^2)\right]+\frac{a^2h^2}{r^2\sin\theta}\right)\ ,
\ee
having used the notation $h_r=\pa_rh$ etc.

Inspection of the first four terms in \re{quadcov} implies that the
functions $h$ and $g$ must tend to {\em constants} for large $r$, by
the requirement of {\em finite energy}. But we see from the third member
of \re{axial} that the requirement that $\phi^{\al}$ be differentiable on
the $z$-axis implies the vanishing of $h$ for $\theta=0,\pi$, idependently
of $r$. From these two conditions it follows that
\be
\label{zax}
\lim_{r\to\infty}h(r,\theta)=0\quad ,
\quad\lim_{r\to\infty}g(r,\theta)=1\ .
\ee

The asymptotic conditions \re{zax} imply the vanishing of the topological
charge
\be
\label{top}
\int\varrho\ d^3x=\eta^3N\int\left[h(gh_{\theta}
-hg_{\theta})\right]_{r=\infty}d\theta\ .
\ee

The topological charge \re{top} will be nonvanishing, namely equal to
$\eta^3N$, if the asymptotic values \re{zax} are replaced by
\be
\label{nontrivial}
\lim_{r\to\infty}h(r,\theta)=\sin\theta\quad ,
\quad\lim_{r\to\infty}g(r,\theta)=\cos\theta\ ,
\ee
which is consistent with the condition of differentaibility on the
$z$-axis, but is not consistent with the finite condition on
\re{quadcov}. This necessitates the replacement of the definition
\re{rho3} by
\bea
\varrho_0&=&\frac{1}{4\pi}\vep_{ijk}\vep^{abc}(\eta^2-|\phi^{\al}|^2)\
\pa_i\phi^a\pa_j\phi^b\pa_k\phi^c\label{Vrho}\\
&=&\frac{1}{4\pi}\vep_{ijk}\vep^{abc}\
\pa_i\ (\eta^2-\frac35|\phi^{\al}|^2)\ \phi^a\pa_j\phi^b\pa_k\phi^c\ .
\label{Vrhodiv}
\eea
The corresponding formula to \re{top} is readily found using \re{Vrhodiv},
and using the new topological charge \re{Vrho}, \re{Vrhodiv}, it turns
out that in the resulting energy density functional, the usual kinetic
term \re{usual} is replaced by the unusual one
\be
\label{unusual}
(\eta^2-|\phi^a|^2)^2|D_i\phi^a|^2\ ,
\ee
reminicent of the $d$ (Euclidean) dimensional $SO(d)$ Higgs
models~\cite{models2,models3,models4} that result from the dimensional
descent of the generalised Yang-Mills~\cite{GYM} (GYM) model on
$\R^d\times S_{4p-d}$. The price for analyticity and topological stability
is the appearance of unconventional kinetic terms \re{unusual}.

\medskip
\noindent
{\bf Acknowledgement}: I am indepted to Burkhard Kleihaus for numerous
illuminating discussions.

\begin{small}

\end{small}
\medskip
\medskip

\end{document}